\documentclass[preprint,12pt]{elsarticle}
\usepackage{graphicx}
\usepackage{amssymb}
\usepackage{amsmath}
\usepackage{subfig,color}
\usepackage{multirow}
\usepackage{colortbl,color}
\usepackage{algorithm}
\usepackage{multirow}
\usepackage{rotating}
\usepackage{amsthm}
\usepackage{hyperref}
\biboptions{comma,sort&compress}
\journal{ }

\begin{document}

\begin{frontmatter}
\title{Analysis on the Invariant Properties of Constitutive Equations of Hydrodynamics in the Transformation between Different Reference Systems}
\author[kaust]{Jun~Li\footnote[1]{e-mail: lijun04@gmail.com\\URL: \url{http://www.researchgate.net/profile/Jun_Li99/}}}
\address[kaust]{Applied Mathematics and Computational Science\\
King Abdullah University of Science and Technology\\Thuwal, Saudi Arabia}

\begin{abstract}
The velocities of the same fluid particle observed in two different reference systems are two different quantities and they are not equal when the two reference systems have translational and rotational movements relative to each other. Thus, the velocity is variant. But, we prove that the divergences of the two different velocities are always equal, which implies that the divergence of velocity is invariant. Additionally, the strain rate tensor and the gradient of temperature are invariant but, the vorticity and gradient of velocity are variant. Only the invariant quantities are employed to construct the constitutive equations used to calculate the stress tensor and heat flux density, which are objective quantities and thus independent of the reference system. Consequently, the forms of constitutive equations keep unchanged when the corresponding governing equations are transformed between different reference systems. Additionally, we prove that the stress is a second-order tensor since its components in different reference systems satisfy the transformation relationship.        
\end{abstract}
\end{frontmatter}
\section{Introduction}\label{s:intro}
We discuss a classical subject that how the constitutive equations of stress tensor and heat flux density should be constructed such that the obtained formulas have unchanged forms when they are applied in different reference systems, including inertial and noninertial systems. Basically, we need to identify which quantities are invariant although they are defined equally but in different reference systems. This question was well posed in \cite{Cao2005}-\cite{Majid2013} and  Chapter 8 of \cite{Zhang1999} proves that the strain rate tensor is invariant but the gradient of velocity is variant. We provide general analyses on the properties of the divergence and gradient of velocity, strain rate tensor, vorticity and, gradient of objective scalar quantity. Additionally, we also prove that the stress satisfies the basic property as a second order tensor, which was posed as a fundamental question in \cite{Wu2005}. 

\section{Convention on the use of notations}\label{s:notations}
In the reference system $s$ having the origin of coordinates at the point $o$, we use $\vec x$ for the vector $\vec {op}$ connecting the point $o$ to an arbitrary spatial point $p$ (i.e. fluid particle moving at the flow velocity). Let $x_i$ be the components (measurements) of $\vec x$ in $s$ and thus $\vec x=x_i\vec e_i$ where $\vec e_{i, (i=1,2,3)}$ are the unit orthogonal vectors of the reference system $s$. Now, we introduce another reference system $s'$ whose origin of coordinates is located at the point $o'$. In general, $s'$ may have translational and rotational movements relative to $s$. We let $\vec X$ denote the vector $\vec {o'p}$ connecting $o'$ to $p$. We use different notations, $\vec x$ and $\vec X$, since they stand for different quantities although they are equal when the positions of $o$ and $o'$ are the same. Let $\vec y$ denote the vector $\vec {oo'}$ connecting $o$ to $o'$ and then we have: 

\begin{equation}\label{eq:define vec y}
\begin{aligned}
     \vec y=\vec x-\vec X.  
\end{aligned}
\end{equation}

For the same vector $\vec x$, which is an objective quantity and thus independent of the reference system, we have different components in different reference systems. Let $x_j'$ be the components of $\vec x$ in $s'$ and then we have:

\begin{equation}\label{eq:vec x}
\begin{aligned}
     \vec x=x_i\vec e_i=x_j'\vec e_j',  
\end{aligned}
\end{equation}
where $\vec e'_{j,(j=1,2,3)}$ are the unit orthogonal vectors of $s'$. Similarly, we have

\begin{equation}\label{eq:vec X}
\begin{aligned}
     \vec X=X_i\vec e_i=X_j'\vec e_j',
\end{aligned}
\end{equation}
and
\begin{equation}\label{eq:vec y}
\begin{aligned}
     \vec y=y_i\vec e_i=y_j'\vec e_j'.
\end{aligned}
\end{equation}

We define the velocity $\vec v$ of point $p$ observed in $s$ as follows:

\begin{equation}\label{eq:vec v}
\begin{aligned}
    \vec v=\dfrac{{\rm d}x_i}{{\rm d}t}\vec e_i,  
\end{aligned}
\end{equation}
where $\dfrac{{\rm d}}{{\rm d}t}$ is the substantial derivative. In contrast, the velocity of point $p$ observed in $s'$ is denoted by $\vec V$ which is computed as

\begin{equation}\label{eq:vec V}
\begin{aligned}
    \vec V=\dfrac{{\rm d}X_j'}{{\rm d}t}\vec e_j'.  
\end{aligned}
\end{equation}
Again, we use two different notations, $\vec v$ and $\vec V$, since they stand for different velocities although they might be equal in some special cases. Note that $\dfrac{{\rm d}x_j'}{{\rm d}t}\vec e_j'$ is the velocity of $p$ relative to $o$ observed in $s'$ and $\dfrac{{\rm d}x_j'}{{\rm d}t}\vec e_j'\ne\vec V$ since $\vec V$ is the velocity of $p$ relative to $o'$ observed in $s'$. Similarly, $\dfrac{{\rm d}X_i}{{\rm d}t}\vec e_i\ne\vec v$. 

In the above definitions, we apply the operation $\dfrac{{\rm d}}{{\rm d}t}$ only to scalar quantities because the notations may become confusing when applying it to vectors. For example, $\dfrac{{\rm d}\vec x}{{\rm d}t}$ usually intends to define $\vec v$ but the assumed calculation according to the second equality of Eq.~\eqref{eq:vec x} may lead to 

\begin{equation}\label{eq:bad use}
\begin{aligned}
    \dfrac{{\rm d}\vec x}{{\rm d}t}=\dfrac{{\rm d} (x_j'\vec e_j')}{{\rm d}t}=\dfrac{{\rm d} x_j'}{{\rm d}t}\vec e_j'+0\ne\vec v,
\end{aligned}
\end{equation}
unless we specify that $\dfrac{{\rm d} \vec e_j'}{{\rm d}t}\ne0$.  But, the consequent problem to make such specification is that we also need to specify $\dfrac{{\rm d} \vec e_i}{{\rm d}t}\ne0$ in the assumed calculation of $\dfrac{{\rm d}\vec X}{{\rm d}t}=\dfrac{{\rm d} (X_i\vec e_i)}{{\rm d}t}$ for $\vec V$. Another reason to avoid using $\dfrac{{\rm d}\vec x}{{\rm d}t}$ and $\dfrac{{\rm d}\vec X}{{\rm d}t}$ to denote $\vec v$ and $\vec V$, respectively, is that $\vec v\ne\vec V$ even if $\vec x\equiv\vec X$ when $s'$ has rotation relative to $s$ though $\vec y=\vec {oo'}\equiv\vec0$ (see Eq.~\eqref{eq:connect V to v}). To make the expression clear, the operation $\dfrac{{\rm d}}{{\rm d}t}$ is applied only to scalar quantities. 

\section{Basic property of tensors}\label{s:tensor property}  
Usually, we call the temperature as 0th-order tensor and, velocity and stress are 1st-order and 2nd-order tensors, respectively. For the same tensor of $n$th-order ($n\ge1$), we have different component expressions because the components (i.e. measurements) depend on the reference system where the tensor is measured. Due to the objective property of tensor, the components for different reference systems must have inherent connection.   

We define the transformation coefficients $\alpha_{ij}$, which depends on the time $t$ in general, between $s$ and $s'$ as follows: 

\begin{equation}\label{eq:alpha_ij}
\begin{aligned}
    \alpha_{ij}=\vec e_i\cdot\vec e_j'.
\end{aligned}
\end{equation}
Thus, we have: 

\begin{equation}\label{eq:ej' to ei}
\begin{aligned}
    \vec e_i=(\vec e_i\cdot\vec e_j')\vec e_j'=\alpha_{ij}\vec e_j',
\end{aligned}
\end{equation}
and 
\begin{equation}\label{eq:ei to ej'}
\begin{aligned}
    \vec e_j'=(\vec e_j'\cdot\vec e_i)\vec e_i=\alpha_{ij}\vec e_i.
\end{aligned}
\end{equation}

Taking $\vec x$ as an example for the discussion of the property of 1st-order tensor and substituting Eq.~\eqref{eq:ej' to ei} into Eq.~\eqref{eq:vec x}, we get: 

\begin{equation}\label{eq:discuss vec x}
\begin{aligned}
    x_j'\vec e_j'=x_i\vec e_i=x_i\alpha_{ij}\vec e_j', 
\end{aligned}
\end{equation}
which implies: 
\begin{equation}\label{eq:property1 of vec x}
\begin{aligned}
    x_j'=x_i\alpha_{ij}.  
\end{aligned}
\end{equation}
Similar derivation based on Eqs.~\eqref{eq:vec x} and \eqref{eq:ei to ej'} shows: 

\begin{equation}\label{eq:property2 of vec x}
\begin{aligned}
    x_i=x_j'\alpha_{ij}.  
\end{aligned}
\end{equation}
Generally, the components of an arbitrary $n$th-order tensor $\mathbf{A}$ ($n\ge1$) satisfy: 

\begin{equation}\label{eq:property2 of A}
\begin{aligned}
    A_{i_1i_2\cdots i_n}=A'_{j_1j_2\cdots j_n}\alpha_{i_1j_1}\alpha_{i_2j_2}\cdots\alpha_{i_nj_n},  
\end{aligned}
\end{equation}
due to Eq.~\eqref{eq:ei to ej'} and 
\begin{equation}\label{eq:A in s and s'}
\begin{aligned}
    \mathbf{A}=A_{i_1i_2\cdots i_n}\vec e_{i_1}\vec e_{i_2}\cdots\vec e_{i_n}=A'_{j_1j_2\cdots j_n}\vec e'_{j_1}\vec e'_{j_2}\cdots\vec e'_{j_n}.  
\end{aligned}
\end{equation}

Additionally, according to Eqs.~\eqref{eq:property1 of vec x} and \eqref{eq:property2 of vec x}, we get: 

\begin{equation}\label{eq:property1 of alpha_ij}
\begin{aligned}
    \alpha_{ij}\alpha_{ik}=\delta_{jk},  
\end{aligned}
\end{equation}
and 
\begin{equation}\label{eq:property2 of alpha_ij}
\begin{aligned}
    \alpha_{ij}\alpha_{kj}=\delta_{ik}.  
\end{aligned}
\end{equation}
where the Kronecker delta is defined as: $\delta_{ij}=1$ for $i=j$ and $\delta_{ij}=0$ for $i\ne j$. 

\section{Quantities with invariant or variant properties}\label{s:invariant and variant}  
As discussed in Section~\ref{s:notations}, we have $x_i\vec e_i=x_i'\vec e_i'$ and usually $x_i\ne x_i'$ for the same objective quantity $\vec x$. Thus, it is pointless to say that $\vec x$ is invariant or variant in the transformation between $s$ and $s'$. The objects of discussion here are not objective quantities but those counterparts which are observed equally but in different reference systems. For example, it makes sense to discuss whether the velocity $\vec v$ of point $p$ observed in $s$ is equal to its velocity $\vec V$ observed in $s'$. Note that $\vec v$ and $\vec V$ are not the same quantity as mentioned before although they might be equal in some special cases. Obviously, the velocity of point $p$ is variant since $\vec v=\vec V$ is not always true. Generally, according to Eqs.~\eqref{eq:define vec y},\eqref{eq:vec v},\eqref{eq:vec V},\eqref{eq:ei to ej'},\eqref{eq:property2 of vec x}, we have:

\begin{equation}\label{eq:connect V to v}
\begin{aligned}
    \vec v&=\dfrac{{\rm d}x_i}{{\rm d}t}\vec e_i \\
              &=\dfrac{{\rm d}(y_i+X_i)}{{\rm d}t}\vec e_i \\
              &=\dfrac{{\rm d}y_i}{{\rm d}t}\vec e_i+\dfrac{{\rm d}X_i}{{\rm d}t}\vec e_i \\ 
              &=\dfrac{{\rm d}y_i}{{\rm d}t}\vec e_i+\dfrac{{\rm d}(\alpha_{ij}X_j')}{{\rm d}t}\vec e_i \\
              &=\dfrac{{\rm d}y_i}{{\rm d}t}\vec e_i+\dfrac{{\rm d}X_j'}{{\rm d}t}\alpha_{ij}\vec e_i+\dfrac{{\rm d}\alpha_{ij}}{{\rm d}t}X_j'\vec e_i \\
              &=\dfrac{{\rm d}y_i}{{\rm d}t}\vec e_i+\dfrac{{\rm d}X_j'}{{\rm d}t}\vec e_j'+\dfrac{{\rm d}\alpha_{ij}}{{\rm d}t}X_j'\vec e_i \\
              &=\dfrac{{\rm d}y_i}{{\rm d}t}\vec e_i+\vec V+\dfrac{{\rm d}\alpha_{ij}}{{\rm d}t}X_j'\vec e_i,
\end{aligned}
\end{equation}
where $\dfrac{{\rm d}y_i}{{\rm d}t}\vec e_i$ and $\dfrac{{\rm d}\alpha_{ij}}{{\rm d}t}X_j'\vec e_i$ correspond to the translational and rotational movements of the reference system $s'$ relative to $s$, respectively. 

We introduce $\vec\omega=\omega_i\vec e_i$ to make the physical meaning of  $\dfrac{{\rm d}\alpha_{ij}}{{\rm d}t}X_j'\vec e_i$ clear and define $\omega_i$ as follows: 

\begin{equation}\label{eq:define omega_i}
\begin{aligned}
    \omega_i=\dfrac{1}{2}\epsilon_{lik}\alpha_{kj}\dfrac{{\rm d}\alpha_{lj}}{{\rm d}t},
\end{aligned}
\end{equation}
where the Levi-Civita symbol is defined as: $\epsilon_{ijk}=0$ if $(i-j)(i-k)(j-k)=0$, $\epsilon_{ijk}=1$ if $(i,j,k)\in\{(1,2,3),(2,3,1),(3,1,2)\}$ and $\epsilon_{ijk}=-1$ if $(i,j,k)\in\{(1,3,2),(3,2,1),(2,1,3)\}$. According to Eqs.~\eqref{eq:property2 of alpha_ij} and \eqref{eq:define omega_i}, we have: 

\begin{equation}\label{eq:apply omega_i}
\begin{aligned}
    \epsilon_{ijk}\omega_j&=\dfrac{1}{2}\epsilon_{ijk}\epsilon_{ljn}\alpha_{nm}\dfrac{{\rm d}\alpha_{lm}}{{\rm d}t} \\
                                         &=\dfrac{1}{2}(\delta_{il}\delta_{kn}-\delta_{in}\delta_{kl})\alpha_{nm}\dfrac{{\rm d}\alpha_{lm}}{{\rm d}t} \\
                                         &=\dfrac{1}{2}(\alpha_{km}\dfrac{{\rm d}\alpha_{im}}{{\rm d}t}-\alpha_{im}\dfrac{{\rm d}\alpha_{km}}{{\rm d}t}) \\
                                         &=\dfrac{1}{2}(2\alpha_{km}\dfrac{{\rm d}\alpha_{im}}{{\rm d}t}-\dfrac{{\rm d}\alpha_{im}\alpha_{km}}{{\rm d}t}) \\
                                         &=\alpha_{km}\dfrac{{\rm d}\alpha_{im}}{{\rm d}t}-\dfrac{1}{2}\dfrac{{\rm d}\delta_{ik}}{{\rm d}t} \\
                                         &=\alpha_{km}\dfrac{{\rm d}\alpha_{im}}{{\rm d}t}.
\end{aligned}
\end{equation}
According to Eqs.~\eqref{eq:property1 of vec x} and \eqref{eq:apply omega_i}, we get: 

\begin{equation}\label{eq:rotational velocity}
\begin{aligned}
     \dfrac{{\rm d}\alpha_{ij}}{{\rm d}t}X_j'\vec e_i&=\dfrac{{\rm d}\alpha_{ij}}{{\rm d}t}\alpha_{kj}X_k\vec e_i \\
                                                                               &=\epsilon_{ijk}\omega_jX_k\vec e_i \\
                                                                               &=\vec\omega\times\vec X,
\end{aligned}
\end{equation}
which shows that $\vec\omega$ defined by Eq.~\eqref{eq:define omega_i} is the rotational speed of $s'$ relative to $s$. 
 
\subsection{Divergence of velocity}\label{ss:divergence}  
We discuss whether $\nabla\cdot\vec v$ is equal to $\nabla'\cdot\vec V$. Note that $y_i$, $y_j'$, $\alpha_{ij}$, $\vec e_i$ and $\vec e_j'$ are independent of $\vec x$ and $\vec X$. According to Eqs.~\eqref{eq:define vec y}, \eqref{eq:alpha_ij}, \eqref{eq:property1 of vec x}, \eqref{eq:property1 of alpha_ij} and \eqref{eq:connect V to v}, we get: 

\begin{equation}\label{eq:divergence}
\begin{aligned}
    \nabla\cdot\vec v&=\vec e_k\dfrac{\partial}{\partial x_k}\cdot(\dfrac{{\rm d}y_i}{{\rm d}t}\vec e_i+\vec V+\dfrac{{\rm d}\alpha_{ij}}{{\rm d}t}X_j'\vec e_i) \\
                                 &=0+\vec e_k\dfrac{\partial V_j'}{\partial x_k}\cdot\vec e_j'+\vec e_k\dfrac{\partial X_j'}{\partial x_k}\dfrac{{\rm d}\alpha_{ij}}{{\rm d}t}\cdot\vec e_i \\
                                 &=\alpha_{kj}\dfrac{\partial X_i'}{\partial x_k}\dfrac{\partial V_j'}{\partial X_i'}+\vec e_k\dfrac{\partial (x_j'-y_j')}{\partial x_k}\dfrac{{\rm d}\alpha_{ij}}{{\rm d}t}\cdot\vec e_i \\
                                 &=\alpha_{kj}\dfrac{\partial (x_i'-y_i')}{\partial x_k}\dfrac{\partial V_j'}{\partial X_i'}+\vec e_k\alpha_{kj}\dfrac{{\rm d}\alpha_{ij}}{{\rm d}t}\cdot\vec e_i-0 \\
                                 &=\alpha_{kj}\alpha_{ki}\dfrac{\partial V_j'}{\partial X_i'}-0+\alpha_{ij}\dfrac{{\rm d}\alpha_{ij}}{{\rm d}t} \\
                                 &=\delta_{ji}\dfrac{\partial V_j'}{\partial X_i'}+\dfrac{1}{2}\dfrac{{\rm d}(\alpha_{ij}\alpha_{ij})}{{\rm d}t} \\
                                 &=\vec e_i'\dfrac{\partial V_j'}{\partial X_i'}\cdot\vec e_j'+\dfrac{1}{2}\dfrac{{\rm d}\delta_{jj}}{{\rm d}t} \\
                                 &=\nabla'\cdot\vec V.
\end{aligned}
\end{equation}
Since $\nabla\cdot\vec v\equiv\nabla'\cdot\vec V$, the divergence of velocity is invariant. 

\subsection{Gradient of objective scalar}\label{ss:scalar gradient}  
We take the temperature $T$ as an example of the objective scalar quantities. The gradients $\nabla T$ and $\nabla' T$ defined in $s$ and $s'$, respectively, are two objective quantities. According to Eqs.~\eqref{eq:define vec y}, \eqref{eq:ei to ej'} and \eqref{eq:property1 of vec x}, we have: 

\begin{equation}\label{eq:scalar gradient}
\begin{aligned}
    \nabla T&=\dfrac{\partial T}{\partial x_i}\vec e_i=\dfrac{\partial X_j'}{\partial x_i}\dfrac{\partial T}{\partial X_j'}\vec e_i=\dfrac{\partial (x_j'-y_j')}{\partial x_i}\dfrac{\partial T}{\partial X_j'}\vec e_i=\alpha_{ij}\dfrac{\partial T}{\partial X_j'}\vec e_i=\dfrac{\partial T}{\partial X_j'}\vec e_j'=\nabla' T.
\end{aligned}
\end{equation}
Since $\nabla T\equiv\nabla'T$, the gradients of temperature and other objective scalar quantities are invariant. 
 
\subsection{Gradient of velocity}\label{ss:velocity gradient}  
According to Eqs.~\eqref{eq:define vec y}, \eqref{eq:ei to ej'}, \eqref{eq:property1 of vec x} and \eqref{eq:connect V to v}, we get: 

\begin{equation}\label{eq:velocity gradient}
\begin{aligned}
    \nabla\vec v&=\vec e_k\dfrac{\partial}{\partial x_k}(\dfrac{{\rm d}y_i}{{\rm d}t}\vec e_i+\vec V+\dfrac{{\rm d}\alpha_{ij}}{{\rm d}t}X_j'\vec e_i) \\
                                 &=0+\dfrac{\partial V_j'}{\partial x_k}\vec e_k\vec e_j'+\dfrac{\partial X_j'}{\partial x_k}\dfrac{{\rm d}\alpha_{ij}}{{\rm d}t}\vec e_k\vec e_i \\
                                 &=\dfrac{\partial X_i'}{\partial x_k}\dfrac{\partial V_j'}{\partial X_i'}\vec e_k\vec e_j'+\dfrac{\partial (x_j'-y_j')}{\partial x_k}\dfrac{{\rm d}\alpha_{ij}}{{\rm d}t}\vec e_k\vec e_i \\
                                 &=\dfrac{\partial (x_i'-y_i')}{\partial x_k}\dfrac{\partial V_j'}{\partial X_i'}\vec e_k\vec e_j'+\alpha_{kj}\dfrac{{\rm d}\alpha_{ij}}{{\rm d}t}\vec e_k\vec e_i-0 \\
                                 &=\alpha_{ki}\dfrac{\partial V_j'}{\partial X_i'}\vec e_k\vec e_j'-0+\alpha_{kj}\dfrac{{\rm d}\alpha_{ij}}{{\rm d}t}\vec e_k\vec e_i \\
                                 &=\dfrac{\partial V_j'}{\partial X_i'}\vec e_i'\vec e_j'+\alpha_{kj}\dfrac{{\rm d}\alpha_{ij}}{{\rm d}t}\vec e_k\vec e_i \\
                                 &=\nabla'\vec V+\alpha_{kj}\dfrac{{\rm d}\alpha_{ij}}{{\rm d}t}\vec e_k\vec e_i.
\end{aligned}
\end{equation}
Since $\nabla\vec v=\nabla'\vec V$ is not always true, the gradient of velocity is variant. 

\subsection{Strain rate tensor}\label{ss:strain rate tensor}  
According to Eqs.~\eqref{eq:property2 of alpha_ij} and \eqref{eq:velocity gradient}, we have: 

\begin{equation}\label{eq:strain rate tensor}
\begin{aligned}
    \dfrac{1}{2}(\nabla\vec v+(\nabla\vec v)^{\rm T})&=\dfrac{1}{2}(\nabla'\vec V+(\nabla'\vec V)^{\rm T}+\alpha_{kj}\dfrac{{\rm d}\alpha_{ij}}{{\rm d}t}\vec e_k\vec e_i+\alpha_{kj}\dfrac{{\rm d}\alpha_{ij}}{{\rm d}t}\vec e_i\vec e_k) \\
                                                                                   &=\dfrac{1}{2}(\nabla'\vec V+(\nabla'\vec V)^{\rm T}+\alpha_{kj}\dfrac{{\rm d}\alpha_{ij}}{{\rm d}t}\vec e_k\vec e_i+\alpha_{ij}\dfrac{{\rm d}\alpha_{kj}}{{\rm d}t}\vec e_k\vec e_i) \\
                                                                                   &=\dfrac{1}{2}(\nabla'\vec V+(\nabla'\vec V)^{\rm T}+\dfrac{{\rm d}(\alpha_{kj}\alpha_{ij})}{{\rm d}t}\vec e_k\vec e_i) \\
                                                                                   &=\dfrac{1}{2}(\nabla'\vec V+(\nabla'\vec V)^{\rm T}+\dfrac{{\rm d}\delta_{ki}}{{\rm d}t}\vec e_k\vec e_i) \\
                                                                                   &=\dfrac{1}{2}(\nabla'\vec V+(\nabla'\vec V)^{\rm T}).
\end{aligned}
\end{equation}
Since $\dfrac{1}{2}(\nabla\vec v+(\nabla\vec v)^{\rm T})\equiv\dfrac{1}{2}(\nabla'\vec V+(\nabla'\vec V)^{\rm T})$, the strain rate tensor is invariant.

\subsection{Vorticity}\label{ss:vorticity}  
According to Eqs.~\eqref{eq:define vec y}, \eqref{eq:ei to ej'}, \eqref{eq:property1 of vec x}, \eqref{eq:connect V to v} and \eqref{eq:rotational velocity}, we have: 

\begin{equation}\label{eq:vorticity}
\begin{aligned}
    \nabla\times\vec v&=\nabla\times(\dfrac{{\rm d}y_i}{{\rm d}t}\vec e_i+\vec V+\omega\times\vec X) \\
                                   &=0+\dfrac{\partial V_i'}{\partial x_j}\vec e_j\times\vec e_i'+\epsilon_{ijk}\dfrac{\partial}{\partial x_j}(\epsilon_{klm}\omega_lX_m)\vec e_i \\
                                   &=\dfrac{\partial X_k'}{\partial x_j}\dfrac{\partial V_i'}{\partial X_k'}\vec e_j\times\vec e_i'+\epsilon_{ijk}\epsilon_{klm}\dfrac{\partial X_m}{\partial x_j}\omega_l\vec e_i \\
                                   &=\dfrac{\partial (x_k'-y_k')}{\partial x_j}\dfrac{\partial V_i'}{\partial X_k'}\vec e_j\times\vec e_i'+(\delta_{il}\delta_{jm}-\delta_{im}\delta_{jl})\dfrac{\partial (x_m-y_m)}{\partial x_j}\omega_l\vec e_i \\
                                   &=\alpha_{jk}\dfrac{\partial V_i'}{\partial X_k'}\vec e_j\times\vec e_i'+(\delta_{il}\delta_{jm}-\delta_{im}\delta_{jl})\delta_{jm}\omega_l\vec e_i \\
                                   &=\dfrac{\partial V_i'}{\partial X_k'}\vec e_k'\times\vec e_i'+3\omega_l\vec e_l-\omega_l\vec e_l \\
                                   &=\nabla'\times\vec V+2\vec\omega. 
\end{aligned}
\end{equation}
Since $\nabla\times\vec v=\nabla'\times\vec V$ is not always true, the vorticity is variant. 

\section{Proof of the stress as a 2nd-order tensor }\label{s:stress tensor}
Usually, we let the component express of stress $\mbox{\boldmath$\tau$}$ have two indexes. It is fine to use this expression as just a notation. But, we should be very careful to relate $\mbox{\boldmath$\tau$}$ to $-p\mathbf{I}+2\mu\mathbf{S}=-p\mathbf{I}+\mu(\nabla\vec v+(\nabla\vec v)^{\rm T})$, where $\mathbf{I}$ is the identity tensor, $p$ the pressure and $\mu$ the dynamic viscosity. The components of $\mathbf{S}$ in different reference systems satisfy Eq.~\eqref{eq:property2 of A} since $\mathbf{S}$ defined above is an objective 2nd-order tensor (\textit{note}: $\dfrac{1}{2}(\nabla'\vec V+(\nabla'\vec V)^{\rm T})$ is another objective 2nd-order tensor although it is equal to $\mathbf{S}$ as proved in Section~\ref{ss:strain rate tensor}). The identity tensor $\mathbf{I}$ also satisfies Eq.~\eqref{eq:property2 of A}. The stress that we are talking about is objective but, we need to prove that it satisfies Eq.~\eqref{eq:property2 of A} as a tensor such that the construction of $\mbox{\boldmath$\tau$}=-p\mathbf{I}+2\mu\mathbf{S}$ is compatible (\textit{note} : can also depend on the divergence of velocity which is invariant as proved in Section~\ref{ss:divergence}). 

For example, we first prove that the vector $\vec x$ satisfies Eq.~\eqref{eq:property2 of A} as a 1st-order tensor. We define $x_i=\vec x\cdot\vec e_i$ and then $(\vec x-x_i\vec e_i)\cdot\vec e_j=x_j-x_j\equiv0$, which implies that $\vec x=x_i\vec e_i$. We also define $x_j'=\vec x\cdot\vec e_j'$ and thus $\vec x=x_j'\vec e_j'$. Now, we have $x_i\vec e_i=\vec x=x_j'\vec e_j'=x_j'\alpha_{ij}\vec e_i$, which proves $x_i=x_j'\alpha_{ij}$ as required by Eq.~\eqref{eq:property2 of A}. The important feature here is that $x_i$ and $x_j'$ are properly defined such that $x_i\vec e_i=x_j'\vec e_j'$ is true.   

For the expression of stress, we define $\tau_{i_1i_2, (i_1=2,i_2=1)}$ as the component at the $y$ direction of the force per unit area exerted by the right side to the left side separated by an imagined plane perpendicular to the $x$ axis. Similar definition applies to $\tau'_{j_1j_2}$ for the observations in $s'$. We can construct two component expressions $\tau_{i_1i_2}\vec e_{i_1}\vec e_{i_2}$ and $\tau'_{j_1j_2}\vec e'_{j_1}\vec e'_{j_2}$ but the previous procedure of proof cannot proceed since we didn't show $\tau_{i_1i_2}\vec e_{i_1}\vec e_{i_2}=\tau'_{j_1j_2}\vec e'_{j_1}\vec e'_{j_2}$ yet. 

We use $\vec f_{\vec n}$ denote the force per unit area exerted by the positive side to the negative side of an imagined plane perpendicular to $\vec n$. For example, $\vec f_{\vec e_1}\cdot\vec e_2=\tau_{21}$. In general, we have $\vec f_{\vec e_{i_2}}\cdot\vec e_{i_1}=\tau_{i_1i_2}$ in $s$ and $\vec f_{\vec e'_{j_2}}\cdot\vec e'_{j_1}=\tau'_{j_1j_2}$ in $s'$. Based on the physical analysis of force balance, we have \cite{Zhang1999}: 
 
\begin{equation}\label{eq:force balance}
\begin{aligned}
    \vec f_{\vec n}=\vec f_{\vec e_1}(\vec n\cdot\vec e_1)+\vec f_{\vec e_2}(\vec n\cdot\vec e_2)+\vec f_{\vec e_3}(\vec n\cdot\vec e_3),
\end{aligned}
\end{equation}
where $\vec n$ is an arbitrary vector. According to Eqs.~\eqref{eq:alpha_ij}, \eqref{eq:ei to ej'} and \eqref{eq:force balance}, we have:  

\begin{equation}\label{eq:tensor property1 of stress}
\begin{aligned}
    \tau'_{j_1j_2}&=\vec f_{\vec e'_{j_2}}\cdot\vec e'_{j_1} \\
                         &=(\vec f_{\vec e_1}(\vec e'_{j_2}\cdot\vec e_1)+\vec f_{\vec e_2}(\vec e'_{j_2}\cdot\vec e_2)+\vec f_{\vec e_3}(\vec e'_{j_2}\cdot\vec e_3))\cdot\vec e'_{j_1} \\
                         &=(\vec f_{\vec e_1}\alpha_{1j_2}+\vec f_{\vec e_2}\alpha_{2j_2}+\vec f_{\vec e_3}\alpha_{3j_2})\cdot\alpha_{i_1j_1}\vec e_{i_1} \\
                         &=(\tau_{i_11}\alpha_{1j_2}+\tau_{i_12}\alpha_{2j_2}+\tau_{i_13}\alpha_{3j_2})\alpha_{i_1j_1} \\
                         &=\tau_{i_1i_2}\alpha_{i_2j_2}\alpha_{i_1j_1},
\end{aligned}
\end{equation}
which implies (substituting Eq.~\eqref{eq:property2 of alpha_ij}): 

\begin{equation}\label{eq:tensor property2 of stress}
\begin{aligned}
    \tau'_{j_1j_2}\alpha_{i_1j_1}\alpha_{i_2j_2}&=\tau_{i_3i_4}\alpha_{i_4j_2}\alpha_{i_3j_1}\alpha_{i_1j_1}\alpha_{i_2j_2} \\
                                                                            &=\tau_{i_3i_4}\delta_{i_4i_2}\delta_{i_3i_1} \\
                                                                            &=\tau_{i_1i_2}.
\end{aligned}
\end{equation}
Eq.~\eqref{eq:tensor property2 of stress} is consistent with the basic property Eq.~\eqref{eq:property2 of A} of tensors and implies $\tau_{i_1i_2}\vec e_{i_1}\vec e_{i_2}=\tau'_{j_1j_2}\vec e'_{j_1}\vec e'_{j_2}$, which are the tensor expressions of the stress in the two reference systems and denoted simply by $\mbox{\boldmath$\tau$}$. Eq.~\eqref{eq:tensor property2 of stress} is true for both fluid and solid systems since Eq.~\eqref{eq:force balance} is valid to both.    

 If we mathematically define $\tau_{i_1i_2}=(\mbox{\boldmath$\tau$}\cdot\vec e_{i_2})\cdot\vec e_{i_1}$ and then $((\mbox{\boldmath$\tau$}-\tau_{i_1i_2}\vec e_{i_1}\vec e_{i_2})\cdot\vec e_{i_4})\cdot\vec e_{i_3}=\tau_{i_3i_4}-\tau_{i_3i_4}\equiv0$, which implies $\mbox{\boldmath$\tau$}=\tau_{i_1i_2}\vec e_{i_1}\vec e_{i_2}$. We also have $\mbox{\boldmath$\tau$}=\tau'_{j_1j_2}\vec e'_{j_1}\vec e'_{j_2}$ where $\tau'_{j_1j_2}$ is defined as $\tau'_{j_1j_2}=(\mbox{\boldmath$\tau$}\cdot\vec e'_{j_2})\cdot\vec e'_{j_1}$. Thus, we have $\tau_{i_1i_2}\vec e_{i_1}\vec e_{i_2}=\mbox{\boldmath$\tau$}=\tau'_{j_1j_2}\vec e'_{j_1}\vec e'_{j_2}$, which implies $\tau_{i_1i_2}=\tau'_{j_1j_2}\alpha_{i_1j_1}\alpha_{i_2j_2}$ according to Eq.~\eqref{eq:ei to ej'}. But, the issue of this derivation is that we don't know the physical meaning of $\tau_{i_1i_2}$ and $\tau'_{j_1j_2}$ which are mathematically defined here. Consequently, the derived results are irrelevant to the stress. Thus, the physical property Eq.~\eqref{eq:force balance} of the stress is the essence which makes the stress as a 2nd-order tensor.   
 
\section{Applications}\label{s:applications}
For an arbitrary vector $\vec b$, we use notations $\dot{\vec b}_{{\rm in} \, s}$ and $\dot{\vec b}_{{\rm in} \, s'}$ for the substantial derivatives of $\vec b$ in $s$ and $s'$, respectively, as follows:   

\begin{equation}\label{eq:time derivative in s}
\begin{aligned}
    \dot{\vec b}_{{\rm in} \, s}=\dfrac{{\rm d} b_i}{{\rm d} t}\vec e_i,
\end{aligned}
\end{equation}
and
\begin{equation}\label{eq:time derivative in s'}
\begin{aligned}
    \dot{\vec b}_{{\rm in} \, s'}=\dfrac{{\rm d} b_j'}{{\rm d} t}\vec e_j'.
\end{aligned}
\end{equation}

We assume that $s$ is an inertial frame of reference and thus the Navier-Stokes momentum equation in $s$ is: 

\begin{equation}\label{eq:N-S in s}
\begin{aligned}
    \rho\dot{\vec v}_{{\rm in} \, s}=-\nabla p+\mu\nabla\cdot(\nabla\vec v+(\nabla\vec v)^{\rm T})+\rho\vec g,
\end{aligned}
\end{equation}
where $\rho$ is the mass density and $\vec g$ the external force per unit mass. In Eq.~\eqref{eq:N-S in s}, we applied the constitutive equation $\mbox{\boldmath$\tau$}=-p\mathbf{I}+\mu(\nabla\vec v+(\nabla\vec v)^{\rm T})$ which is valid in $s$ according to the experimental observations conducted in the inertial frame of reference. According to Section~\ref{ss:strain rate tensor} and even without additional experimental verifications in the noninertial system $s'$, we have $\mbox{\boldmath$\tau$}=-p\mathbf{I}+\mu(\nabla'\vec V+(\nabla'\vec V)^{\rm T})$ which implies that the constitutive equation of stress tensor has an unchanged form when it is applied in different reference systems. Additionally, we have $\nabla p\equiv\nabla' p$ as discussed in Section~\ref{ss:scalar gradient} and 

\begin{equation}\label{eq:gradient of strain rate tensor}
\begin{aligned}
    \nabla\cdot(\nabla'\vec V+(\nabla'\vec V)^{\rm T})&=\vec e_k\dfrac{\partial}{\partial x_k}\cdot(\nabla'\vec V+(\nabla'\vec V)^{\rm T}) \\
                  &=\vec e_k\dfrac{\partial X_l'}{\partial x_k}\dfrac{\partial}{\partial X_l'}\cdot(\nabla'\vec V+(\nabla'\vec V)^{\rm T}) \\
                  &=\vec e_k\dfrac{\partial (x_l'-y_l')}{\partial x_k}\dfrac{\partial}{\partial X_l'}\cdot(\nabla'\vec V+(\nabla'\vec V)^{\rm T}) \\ 
                  &=\vec e_k\alpha_{kl}\dfrac{\partial}{\partial X_l'}\cdot(\nabla'\vec V+(\nabla'\vec V)^{\rm T}) \\
                  &=\vec e_l'\dfrac{\partial}{\partial X_l'}\cdot(\nabla'\vec V+(\nabla'\vec V)^{\rm T}) \\
                  &=\nabla'\cdot(\nabla'\vec V+(\nabla'\vec V)^{\rm T}),
\end{aligned}
\end{equation}
where we submitted Eqs.~\eqref{eq:define vec y}, \eqref{eq:ei to ej'} and \eqref{eq:property1 of vec x}. Thus, we can rewrite Eq.~\eqref{eq:N-S in s} into: 

\begin{equation}\label{eq:N-S in s new}
\begin{aligned}
    \rho\dot{\vec v}_{{\rm in} \, s}=-\nabla' p+\mu\nabla'\cdot(\nabla'\vec V+(\nabla'\vec V)^{\rm T})+\rho\vec g.
\end{aligned}
\end{equation}

Note that $\dot{\vec v}_{{\rm in} \, s}$ is the acceleration which is measurable in $s$ but not in $s'$. To get the momentum equation for $s'$, we need to make sure that all of the quantities in the equation are either measurable in $s'$ or the properties of $s'$ (see Eq.~\eqref{eq:accelerations}). Those properties are measured in $s$ and independent of $\vec X$ since they are related to the whole reference system of $s'$. According to Eqs.~\eqref{eq:vec V}, \eqref{eq:ei to ej'}, \eqref{eq:property1 of vec x}, \eqref{eq:property2 of vec x}, \eqref{eq:connect V to v}, \eqref{eq:apply omega_i}, \eqref{eq:rotational velocity}, \eqref{eq:time derivative in s} and \eqref{eq:time derivative in s'}, we have the connection between the accelerations of particle $p$ observed in $s$ and $s'$, respectively: 

\begin{equation}\label{eq:accelerations} 
\begin{aligned}
    \dot{\vec v}_{{\rm in} \, s}&=\dfrac{{\rm d}^2y_i}{{\rm d}t}\vec e_i+\dfrac{{\rm d}V_i}{{\rm d}t}\vec e_i+\dfrac{{\rm d}(\epsilon_{ijk}\omega_jX_k)}{{\rm d}t}\vec e_i \\
                                              &=\dfrac{{\rm d}^2y_i}{{\rm d}t}\vec e_i+\dfrac{{\rm d}(\alpha_{ij}V_j')}{{\rm d}t}\vec e_i+\epsilon_{ijk}\dfrac{{\rm d}\omega_j}{{\rm d}t}X_k\vec e_i+\epsilon_{ijk}\omega_j\dfrac{{\rm d}X_k}{{\rm d}t}\vec e_i \\
                                              &=\dfrac{{\rm d}^2y_i}{{\rm d}t}\vec e_i+\dfrac{{\rm d}V_j'}{{\rm d}t}\alpha_{ij}\vec e_i+\dfrac{{\rm d}\alpha_{ij}}{{\rm d}t}V_j'\vec e_i+\dot{\vec\omega}_{{\rm in} \, s}\times\vec X+\epsilon_{ijk}\omega_j\dfrac{{\rm d}(\alpha_{kl}X_l')}{{\rm d}t}\vec e_i \\
                                              &=\dfrac{{\rm d}^2y_i}{{\rm d}t}\vec e_i+\dfrac{{\rm d}V_j'}{{\rm d}t}\vec e_j'+\dfrac{{\rm d}\alpha_{ij}}{{\rm d}t}\alpha_{kj}V_k\vec e_i+\dot{\vec\omega}_{{\rm in} \, s}\times\vec X+\epsilon_{ijk}\omega_j\dfrac{{\rm d}(\alpha_{kl}X_l')}{{\rm d}t}\vec e_i \\
                                              &=\dfrac{{\rm d}^2y_i}{{\rm d}t}\vec e_i+\dot{\vec V}_{{\rm in} \, s'}+\epsilon_{ijk}\omega_jV_k\vec e_i+\dot{\vec\omega}_{{\rm in} \, s}\times\vec X+\epsilon_{ijk}\omega_j\dfrac{{\rm d}(\alpha_{kl}X_l')}{{\rm d}t}\vec e_i \\
                                              &=\dfrac{{\rm d}^2y_i}{{\rm d}t}\vec e_i+\dot{\vec V}_{{\rm in} \, s'}+\vec\omega\times\vec V+\dot{\vec\omega}_{{\rm in} \, s}\times\vec X+\epsilon_{ijk}\omega_j(\dfrac{{\rm d}\alpha_{kl}}{{\rm d}t}X_l'+\alpha_{kl}\dfrac{{\rm d}X_l'}{{\rm d}t})\vec e_i \\
                                              &=\dfrac{{\rm d}^2y_i}{{\rm d}t}\vec e_i+\dot{\vec V}_{{\rm in} \, s'}+\vec\omega\times\vec V+\dot{\vec\omega}_{{\rm in} \, s}\times\vec X+\epsilon_{ijk}\omega_j(\dfrac{{\rm d}\alpha_{kl}}{{\rm d}t}\alpha_{ml}X_m+\alpha_{kl}V_l')\vec e_i \\
                                              &=\dfrac{{\rm d}^2y_i}{{\rm d}t}\vec e_i+\dot{\vec V}_{{\rm in} \, s'}+\vec\omega\times\vec V+\dot{\vec\omega}_{{\rm in} \, s}\times\vec X+\epsilon_{ijk}\omega_j(\epsilon_{klm}\omega_lX_m+V_k)\vec e_i \\
                                              &=\dfrac{{\rm d}^2y_i}{{\rm d}t}\vec e_i+\dot{\vec V}_{{\rm in} \, s'}+\vec\omega\times\vec V+\dot{\vec\omega}_{{\rm in} \, s}\times\vec X+\omega\times(\omega\times\vec X)+\omega\times\vec V \\
                                              &=\dfrac{{\rm d}^2y_i}{{\rm d}t}\vec e_i+\dot{\vec V}_{{\rm in} \, s'}+2\vec\omega\times\vec V+\dot{\vec\omega}_{{\rm in} \, s}\times\vec X+\omega\times(\omega\times\vec X).
\end{aligned}
\end{equation}
Substituting Eq.~\eqref{eq:accelerations} into \eqref{eq:N-S in s new}, we get the momentum equation which is valid in an arbitrary noninertial system $s'$. Similar derivation can be applied to the transformation of the energy equation.   


\end{document}